\documentclass[11pt,a4paper]{article}
\usepackage{flafter}
\usepackage[dvips]{graphicx}
\usepackage{latexsym}
\usepackage[fleqn]{amsmath}
\usepackage{algorithmic,algorithm,amsfonts,amssymb}

\newtheorem{theorem}{Theorem}
\newtheorem{lemma}{Lemma}

\newcommand{\CALL}[2]{\mbox{#1}(#2)}
\newcommand{\TO}{~\mbox{\bf to}~}
\newcommand{\KEMBALI}{\mbox{\bf return}~}
\newcommand{\GETS}{\leftarrow}

\renewcommand{\algorithmiccomment}[1]{\hfill{//{\it #1}}}

\def\squarebox#1{\hbox to #1{\hfill\vbox to #1{\vfill}}}
\def\qed{\hspace*{\fill}
        \vbox{\hrule\hbox{\vrule\squarebox{.667em}\vrule}\hrule}\smallskip}
\newenvironment{proof}{\begin{trivlist}
  \item[\hspace{\labelsep}{\em\noindent Proof.~}]
  }{\qed\end{trivlist}}

\begin{document}

\title{Robust Quantum Algorithms for Oracle Identification}

\author{
Andris Ambainis$^{1}$$\qquad$
Kazuo Iwama$^{2}$$\qquad$
Akinori Kawachi$^{3}$\\
Rudy Raymond$^{2}$$\quad$
Shigeru Yamashita$^{4}$\\
\\
$^{1}$Department of Combinatorics and Optimization, University of Waterloo\\
{\tt ambainis@math.uwaterloo.ca}\\
$^{2}$Graduate School of Informatics,  Kyoto University\\
{\tt \{iwama,raymond\}@kuis.kyoto-u.ac.jp}\\
$^{3}$ Graduate School of Information Science and Engineering,
Tokyo Institute of Technology\\ 
{\tt kawachi@is.titech.ac.jp}\\
$^{4}$Graduate School of Information Science, Nara Institute of Science 
and Technology\\ 
{\tt ger@is.naist.jp}
}
\date{}

\maketitle

\begin{abstract}
The oracle identification problem (OIP) was introduced by Ambainis
et al.~\cite{AIKMRY04}. It is given as a set $S$ of $M$ oracles
and a blackbox oracle $f$. Our task is to figure out which oracle in $S$
is equal to the blackbox $f$ by making queries to $f$. OIP includes
several problems such as the Grover Search as special cases. In this 
paper, we improve the algorithms in \cite{AIKMRY04} by providing  a 
mostly optimal upper bound of query complexity for this problem: ($i$) 
For any oracle set $S$ such that $|S| \le 2^{N^d}$  
($d < 1$), we design an algorithm whose query complexity is
$O(\sqrt{N\log{M}/\log{N}})$, matching the lower bound proved in 
\cite{AIKMRY04}. ($ii$) Our algorithm also works for the range between  
$2^{N^d}$ and $2^{N/\log{N}}$ (where the bound becomes $O(N)$), but the 
gap between the upper and lower bounds worsens gradually. ($iii$) Our 
algorithm is robust, namely, it exhibits the same performance (up to 
a constant factor) against the noisy oracles as also shown in 
the literatures \cite{AC02,BNRW03,HMW03} for special cases of OIP. 
\end{abstract}
\hspace{1cm}{\bf keywords:~}quantum computing, query complexity and algorithmic 
learning theory

\section{Introduction}

We study the following problem, called the Oracle Identification 
Problem (OIP): Given a hidden $N$-bit vector $f=(a_1,\ldots,a_N) \in 
\{0,1\}^{N}$, called an {\it oracle}, and a {\it candidate set} $S  
\subseteq \{0,1\}^N$, OIP requires us to find which oracle in $S$ is 
equal to $f$. 
OIP has been especially popular since the emergence of quantum 
computation, e.g., \cite{BV97, BBBGL98, BBHT96, BNRW03, CK97, HMW03}. For 
example, suppose that we set $S = \{(a_1,\ldots,a_N)|  
\mbox{~exactly one~} a_i = 1\}$. 
Then this OIP is essentially the same as Grover search \cite{Gro96}. 
In \cite{AIKMRY04}, Ambainis et al. extended 
the problem to a general $S$. They proved that the total cost of {\it 
any} OIP with $|S| = N$ is $O(\sqrt{N})$, which is optimal within a 
constant factor since this includes the Grover search as a special case 
and for the latter an $\Omega(\sqrt{N})$ lower bound is known (e.g., \cite{BBHT96}). For a  
larger $S$, they obtain nontrivial upper and lower bounds, 
$O(\sqrt{N\log{M}\log{N}}\log{\log{M}})$ and $\Omega(\sqrt{N 
\log{M}/\log{N}})$, respectively, but unfortunately, there is a fairly 
large gap between them.

{\bf Our Result.~} 
Let $M = |S|$. ($i$) If $M \le 2^{N^d}$ for a 
constant $d$ ($< 1$), then the cost of our new algorithm is $O(\sqrt{N 
\log{M}/\log{N}})$ which matches the lower bound obtained in 
\cite{AIKMRY04}. (Previously we have an optimal upper bound only for $M 
= N$). ($ii$) For the range between $2^{N^d}$ and $2^{N/\log{N}}$, our algorithm 
works without any modification and the (gradually growing) gap to the lower bound is at most a factor of 
$O(\sqrt{\log{N}\log\log{N}})$. ($iii$) Our algorithm is robust, namely, it exhibits the 
same performance (up to a constant factor) against the {\em noisy\/} 
oracles as shown in the literatures \cite{AC02,BNRW03,HMW03} for special 
cases of OIP.  

Our algorithms use two operations: ($i$) The first one 
is a simple query ({\it S-query}) to the hidden oracle, i.e., to obtain 
the value ($0$ or $1$) of $a_{i}$ by specifying the $\log{N}$-bit index $i$. The 
cost for this query is one per each. ($ii$) The second one is called 
a {\it G-query} to the oracle: By specifying a set 
$T=\{i_1,\ldots,i_r\}$ of indices, we can obtain, if any, an index $i_j \in T$ s.t. 
$a_{i_j}=1$ and nill otherwise. If there are two or more such $i_j$'s 
then one of them is chosen at random. The cost for this query is 
$O(\sqrt{|T|/K})$ where $K = \left| \{i_j|~i_j \in T \mbox{~and~} 
a_{i_j}=1\}\right|+1$. This query is stochastic, i.e., the 
answer is correct with a constant probability. Obviously our goal is to 
minimize the cost for solving the OIP with a constant success 
probability. Note that we incur the cost for only S- 
and G-queries (i.e., the cost for any other computation is zero), and it 
turns out that our query model is equivalent to the standard  
query complexity one, e.g., \cite{BBCMW98}.      

S-queries are standard and may not need any explanation. G-queries 
are, as one can see, the Grover Search themselves. So, they cannot be 
implemented in the framework of classical computation, and hence our paper is 
definitely a quantum paper. However, if we use the two 
queries as blackbox subroutines and follow the above complexity 
measure, then our algorithm design will be completely classical. Now it 
is important to observe the "efficiency" of G-queries. Since its cost is  
sublinear in $|T|$, our general idea is that it is more cost-effective 
to use them for a larger $T$. For example, the cost for a single 
G-query for $|T| = L$ is less than the total cost of three G-queries for $|T'| = 
L/3$. However, it is also true that the former is less informative since 
it gives us only one bit-position in $T$ which has value one, while the latter 
gives us three. Thus, as one would expect, selecting the size of $T$ is 
a key issue when using G-queries.

As mentioned earlier, if we use the two queries as blackbox subroutines 
together with their cost rule, then any knowledge about quantum 
computation is not needed in the design and analysis of our algorithms. Since $S$ is 
a set of $M$~ $0/1$-vectors of length $N$, it is naturally given as a $0/1$ matrix 
$Z$ of $N$ columns and $M$ rows. For a given $Z$, our basic strategy is 
quite simple: if there is a column which includes a balanced number of 
$0$'s and $1$'s, then we ask the value of the oracle at that position by 
using an S-query. This reduces the number of candidates by a constant 
factor. Otherwise, i.e., if every column has, say, a small fraction of 
$1$'s, then S-queries may seldom reduce the candidates. In such a situation, the 
idea is that it is better to use a G-query by 
selecting a certain number of columns in $T$ than repeating S-queries. 
In order to optimize this strategy, our new algorithm controls the size of $T$ very  
carefully. This contrasts with the previous method~\cite{AIKMRY04} that uses G-queries 
always with $T=\{1,\ldots,N\}$  



{\bf Previous Work.~}Suppose that we wish to solve some problem over 
input data of $N$ bits. Presumably, we need all the values of these $N$ 
bits to obtain a correct answer, which in turn requires $N$ (simple) 
queries to the data. In a certain situation, we do not need all the 
values, which allows us to design a variety of sublinear-time (classical) 
algorithms, e.g., \cite{CLM03, GGR96, KS03}. This is also true when the input 
is given with some premise, for which giving a candidate set as in this paper is 
the most general method. Quickly approaching to the hidden data using 
the premise information is the basis of algorithmic learning theory. In 
fact, Atici et al. in \cite{AS04} independently use techniques similar 
to ours in the context of quantum learning theory. One of their  
results, which states the existence of a quantum algorithm for learning 
a concept class $S$ whose parameter is $\gamma_S$ with 
$O(\log{|S|}\log{\log{|S|}}/\sqrt{\gamma_S})$ queries, almost establishes a conjecture 
of $O(\log{|S|}/\sqrt{\gamma_S})$ queries in \cite{Hunz03}. 

Recall that our complexity measure is the (quantum) query complexity, 
which has been intensively studied as a central issue of quantum 
computation. The most remarkable result is  due to Grover \cite{Gro96}, 
which provided a number of applications and  extensions, e.g., \cite{BBBGL98, BBHT96,
CK97}. Recently quite many results on efficient quantum
algorithms are shown by "sophisticated" ways of using the Grover
Search. (Our present paper is also in this category.) Brassard et 
al.~\cite{BHMT00} showed a quantum counting algorithm that gives an 
approximate counting method by combining the Grover Search with the 
quantum Fourier transformation.  Quantum algorithms for the claw-finding 
and the element distinctness problems given by Buhrman~et 
al.~\cite{BDHHMSW01} also exploited classical random  and sorting 
methods with the Grover Search. (Ambainis~\cite{Amb03b} developed an 
optimal quantum algorithm with  $O(N^{2/3})$ queries for 
element distinctness problem, which makes use of the quantum walk and
matches to the lower bounds shown by Shi \cite{Shi02}.) 
Aaronson et al.~\cite{AA03} constructed quantum search algorithms for 
spatial regions by combining the Grover Search with the
divided-and-conquer method. Magniez et al. \cite{MSS05}
showed efficient quantum algorithms  to find a triangle in a given
graph by using combinatorial techniques  with the Grover
Search. D{\"u}rr et al. \cite{DHHM04} also
investigated  quantum query complexity of several graph-theoretic
problems. In particular, they exploited the Grover Search on some data
structures of graphs for their upper bounds.  

Recently, two papers, by H{\o}yer et al.~\cite{HMW03} and Buhrman 
et al.~\cite{BNRW03}, raised the question of how to cope with
``imperfect'' oracles for the quantum case using  
the following model: The oracle returns, for the query to bit $a_i$, a 
quantum pure state from which we can measure the correct value of
$a_i$ with a constant probability. This noise model naturally fits the
motivation that a similar mechanism should apply when we use
bounded-error quantum subroutines. In \cite{HMW03} H{\o}yer et~al. gave
a quantum algorithm that robustly computes the Grover's problem with
$O(\sqrt{N})$ queries, which is only a constant factor worse than the
noiseless case. Buhrman et al.~\cite{BNRW03} also gave a robust
quantum algorithm to output all the $N$ bits by using $O(N)$
queries. This obviously implies that $O(N)$ queries are enough to
compute the parity of the $N$ bits, which contrasts with the classical
$\Omega(N\log N)$ lower bound given in~\cite{FRPU94}. 
Thus, robust quantum computation does not need a serious overhead at
least for several important problems, including the OIP discussed in
this paper. 

\section{S-queries, G-queries and Robustness}
Recall that an instance of OIP is given as a set $S=\{f_1,\ldots,f_M\}$ 
of oracles, each $f_i = (f_i(1),\ldots,f_i(N)) \in \{0,1\}^N$, and a 
hidden oracle $f \in S$ which is not known in advance. We are asked to 
find the index $i$ such that $f = f_i$. We can access the hidden oracle 
$f$ through a unitary transformation $U_f$, which is referred to as an 
{\it oracle call}, such that       
$$
U_{f}\left|x\right>\left|0\right> =
\left|x\right>\left|f(x)\right>, 
$$
where $1 \le x \le N$ denotes the bit-position of $f$ whose value ($0$ 
or $1$) we wish to know. This bit-position might be a superposition of 
two or more bit-positions, i.e., $\sum_i \alpha_i \left|x_i\right>$. Then the 
result of the oracle call is also a superposition, i.e., $\sum_i \alpha_i 
\left|x_i\right>\left|f(x_i)\right>$. The query complexity counts the 
number of oracle calls being necessary to obtain a correct answer $i$ 
with a constant probability.       

In this paper we will not use oracle calls directly but through two 
subroutines, S-queries and G-queries. (Both can be viewed as classical 
subroutines when used.) An S-query, $\mbox{SQ}(i)$, is simply a 
single oracle call with the index $i$ plus observation. It returns 
$f(i)$ with probability one and its query complexity is obviously one. A 
G-query, $\mbox{GQ}(T)$, where $T \subseteq \{1,\ldots,N\}$, returns $1 
\le i \le N$ such that $i \in T$ and $f(i) = 1$ if such $i$ exists and 
nill otherwise. We admit an error, namely, the answer may be incorrect 
but should be correct with a constant probability, say, $2/3$. Although 
details are omitted, it is easy to see that $\mbox{GQ}(T)$ can be 
implemented by applying Grover Search only to the selected positions $T$. 
Its query complexity is given by the  
following lemma.            
\begin{lemma}[\cite{BHMT00}]
$\mbox{GQ}(T)$ needs $O(\sqrt{|T|/K})$ oracle calls, where $K = \left| 
\{j|~j \in T \mbox{~and~} f(j)=1\}\right|+1$.   
\end{lemma}

If $f$ is a {\it noisy oracle}, then its unitary transformation is given 
as follows \cite{AC02}:  
$$
\tilde{U}_{f}\left|x\right>\left|0\right>\left|0\right> =
\sqrt{p_x}\left|x\right>\left|\phi_{x}\right>\left|f(x)\right> +
\sqrt{1 - p_x}\left|x\right>\left|\psi_{x}\right>\left|\neg{f(x)}\right>, 
$$
where $2/3 \le p_x \le 1$, $\left|\phi_{x}\right>$ and 
$\left|\psi_{x}\right>$ (the states of working registers) may depend on 
$x$. As before $\left|x\right>$ (and hence the result also) may be a 
superposition of bit-positions. Since an oracle call itself includes an 
error, an S-query should also be stochastic. $\tilde{\mbox{SQ}}(i)$ 
returns $f(i)$ with probability at least $2/3$ (and $\neg{f(i)}$ with at 
most $1/3$). G-queries, $\tilde{\mbox{GQ}}(T)$, are already stochastic, 
i.e., succeed to find an answer with probability at least $2/3$ if 
there exists one, and they do not need modification.       

\begin{lemma}[\cite{HMW03}]\label{noisymtgs}
Let $K$ and $T$ be as before. Then $\tilde{\mbox{GQ}}(T)$ needs $O(\sqrt{|T|/K})$ 
noisy oracle calls.   
\end{lemma}

In this paper our oracle mode is almost always noisy. Therefore we 
simply use the notation $\mbox{SQ}$ and $\mbox{GQ}$ instead of $\tilde{\mbox{SQ}}$ 
and $\tilde{\mbox{GQ}}$.    

\section{Algorithms for Small Candidate Sets}

\subsection{Overview of the Algorithm}
Recall that the candidate set $S$ ($|S| = M)$ is given as an $M\times N$ 
matrix $Z$. Before we give our main result in the next section, we 
discuss the case that $Z$ is small, i.e., $M = \mbox{poly}(N)$ in this 
section, which we need in the main algorithm and also will be nice to 
understand the basic idea. Since our goal is to find a single row from the  
$M$ ones, a natural strategy is to reduce the number of candidate rows (a 
subset of rows denoted by $S$) step by step. This can be done easily if 
there is a column, say, $j$ which is ``balanced," i.e., which has an 
approximately equal number of $0$'s and $1$'s in $Z(S)$, where $Z(S)$  
denotes the matrix obtained from $Z$ by deleting all rows not in $S$. 
Then by asking the value of $f(j)$ by an $\mbox{SQ}(j)$, we can reduce 
the {\it size} of $S$ (i.e., the number of oracle candidates) by a constant factor. Suppose 
otherwise, that there are no such good columns in $Z(S)$. Then we gather 
a certain number of columns such that the set $T$ of these columns is 
``balanced," namely, such that the number of rows which has $1$ 
somewhere in $T$ is a constant fraction of $|S|$. (See Fig.~1 where the 
columns in $T$ are shifted to the left.) Now we execute $\mbox{GQ}(T)$ 
and we can reduce the size of $S$ by a constant fraction according to 
whether $\mbox{GQ}(T)$ returns nill ($S$ is reduced to $S_2$ in Fig.~1) 
or not ($S$ is reduced to $S_1$ in Fig.~1). Then we move to the next 
iteration until $|S|$ becomes one.

The merit of using $\mbox{GQ}(T)$ is obvious since it needs at most 
$O(\sqrt{|T|})$ queries while we may need roughly $|T|$  
queries if asking each position by S-queries. Even so, if $|T|$ is too 
large, we cannot tolerate the cost for $\mbox{GQ}(T)$. So, the key issue 
here is to set a carefully chosen upper bound for the size of $T$. If we 
can select $T$ within this upper bound, then we are happy. Otherwise, we 
just give up constructing $T$ and use another strategy which takes 
advantage of the sparseness of the current matrix $Z(S)$. (Obviously 
$Z(S)$ is sparse since we could not select a $T$ of small size.)  

It should be also noted that in each iteration the matrix $Z(S)$ should
be {\em one-sensitive}, namely the number of 1's is less than or equal to
the number of  0's in every column. (The reason is obvious since it
does not make sense to try to find 1 if almost all entries are 1.) For
this purpose we implicitly apply the {\it column-flipping} procedure in each
iteration. Suppose that some column, say $j$, of $Z(S)$ has more 1's than
0's. Then this procedure ``flips" the value of $f(j)$ by adding an
extra circuit to the oracle (but without any oracle call). Let this 
oracle be $\overline{f(j)}$ and  $\overline{Z(S(j))}$ be the matrix 
obtained by flipping the column $j$ of $Z(S)$. Then obviously $f \in S$ 
iff the matrix $\overline{Z(S(j))}$ contains the row $\overline{f(j)}$, 
i.e., the problem does not change essentially. Note that the 
column-flipping is the same as that in \cite{AIKMRY04}, where the OIP matrix 
was written as a $N \times M$ (number of columns $\times$ number of rows) 
$0$-$1$ matrix instead of the more common $M \times N$ one. 
\begin{figure}[h]
\begin{minipage}{.4\linewidth}
\begin{center}
\resizebox{5.5cm}{!}{ 
\includegraphics*{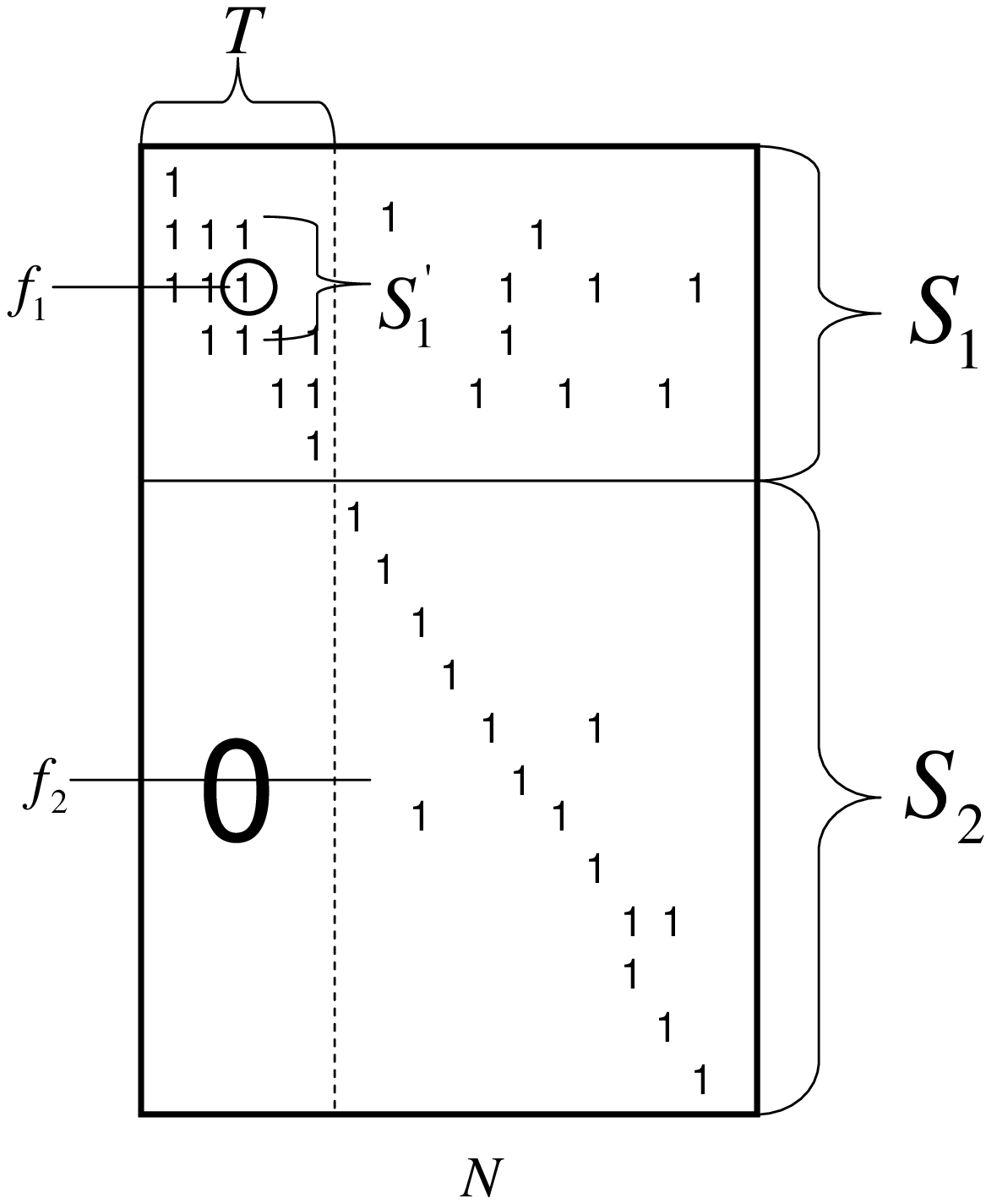} 
}
\end{center}
\caption{Reducing the candidate set by G-queries $\mbox{GQ}(T)$ on the 
column set $T$}
\end{minipage}
\begin{minipage}{.5\linewidth}
\begin{center}
\resizebox{7.5cm}{!}{ 
\includegraphics*{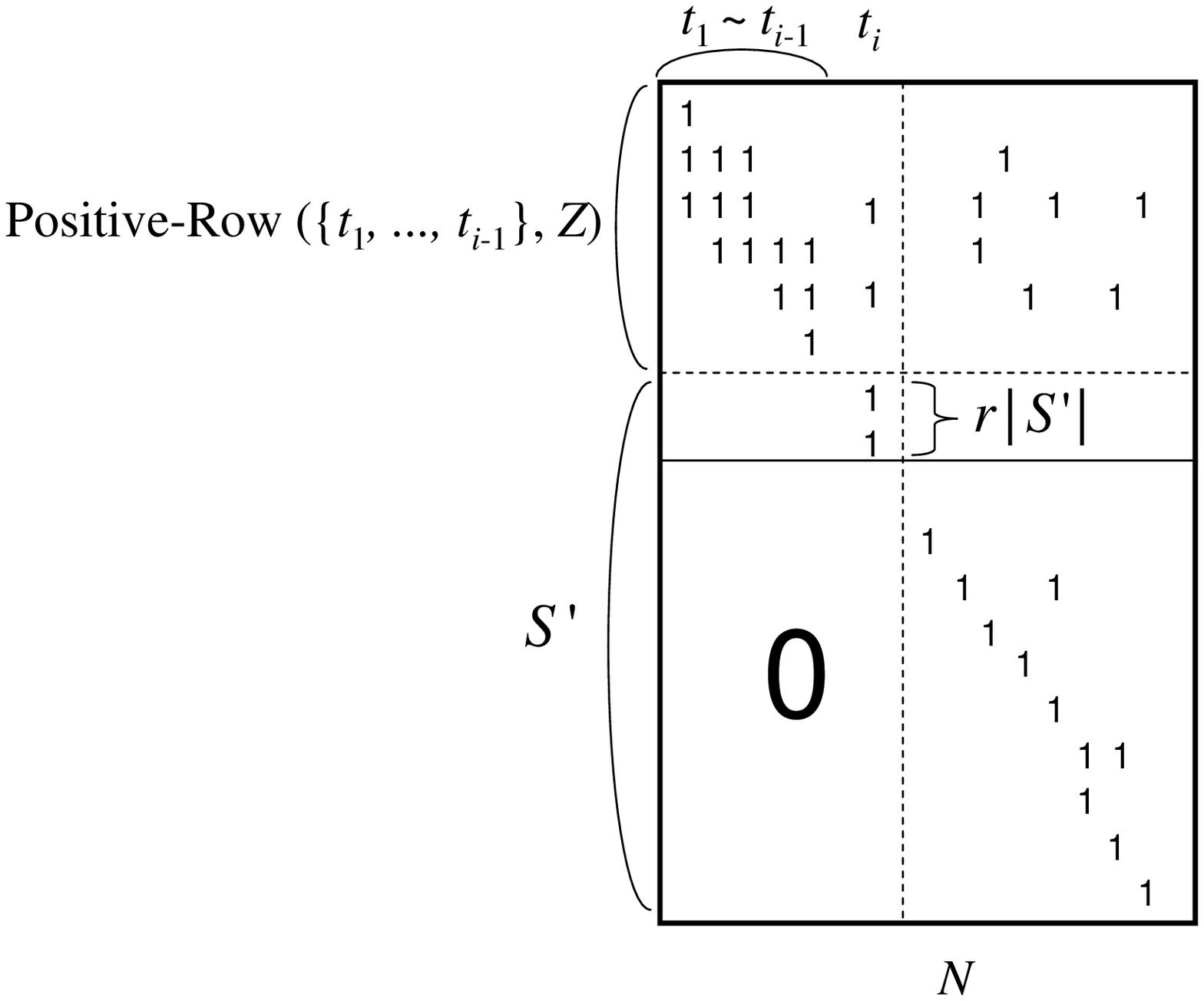} 
}
\end{center}
\caption{Constructing the column set $T$ by RowCover($S,r$)}
\end{minipage}
\end{figure}

\subsection{Procedure RowReduction($T,l$) for Reducing Oracles Candidates}
This procedure narrows $S$ in each iteration, where $T$ is a set of 
columns and $l$ is an integer $\ge 1$ necessary for error control. 
See Procedure 1 for its pseudocode. \underline{Case $1$}: If $f$ has 
one or more $1$'s in $T$ like $f_1$ in Fig.~1, then $k = \mbox{GQ}(T)$ gives 
us one of the positions of these $1$'s, say the circled one in the 
figure. The procedure returns with the set $S_1'$ of rows in the figure, 
i.e., the rows having a $1$ in the position selected by the $\mbox{GQ}(T)$.      
\underline{Case $2$}: If $f$ has no $1$'s in $T$ like $f_2$ in the figure, 
then $k=\mbox{~nill}$ (i.e., $\mbox{GQ}(T)$ correctly answered). Even if 
$k\neq \mbox{~nill}$ ($\mbox{GQ}(T)$ failed) then 
$\CALL{Majority}{k,l,f}$, i.e., the majority of $60l$ samples of $f(k)$, 
is $0$ with high probability regardless of the value of $k$. Therefore 
the procedure returns with the set $S_2$ of rows, i.e., the rows having 
no $1$'s in $T$. The parameter $l$ guarantees the success probability of 
this procedure as follows.      

\begin{lemma}\label{rowreduct}
The success probability and the number of oracle calls in 
$\CALL{RowReduction}{T,l}$ are $1-O(l/3^{l})$ and $l(O(\sqrt{|T|}) + l)$, 
 respectively.  
\end{lemma}
\begin{proof}
In each repetition, we need $O(\sqrt{|T|})$ oracle calls for the 
G-queries and $O(l)$ calls (S-queries) for $\CALL{Majority}{k,l,f}$. Thus 
the total number of calls is $l(O(\sqrt{|T|}) + l)$. For the success 
probability, let us first consider Case $1$ above. Since the G-queries 
are repeated up to $l$ times, the probability that all tries fail (i.e., 
the next $\mbox{Majority}=0$) is $1/3^{l}$. When it succeeds, the 
following $\mbox{Majority}$ fails with probability  
$1/3^{l}$ also (Here, the number of samples ($=60l$) for majority is set 
appropriately so that the error probability is at most $1/3^{l}$ by the 
Chernoff bound). Hence the total failure probability is at most 
$O(1/3^{l})$. In Case $2$, since $\mbox{Majority}$ fails with 
probability $O(1/3^{l})$ in each iteration, the total probability of 
failure is at most $O(l/3^l)$.   
\end{proof}

\subsection{Procedure RowCover($S,r$) for Collecting Position of Queries}
As mentioned in Sec.~3.1, we need to make a set $T$ of 
columns being balanced as a whole. This procedure is used for this 
purpose where $Z(S)$ is the current matrix and $0 < r \le 1$ controls the 
size of $T$. See Procedure 2 for its pseudocode. As shown in Fig.~2, 
the procedure adds columns $t_1,t_2,\ldots, $ to $T$ as long as a new 
addition $t_i$ increases the number of covered rows 
($=|\CALL{PositiveRow}{T,Z}|$) by a factor of $r$ or until the number of 
covered rows becomes $|S|/4$. We say that RowCover  
{\it succeeds} if it finishes with $S'$ such that $|S'| \le 
\frac{3|S|}{4}$ and {\it fails} otherwise. Suppose that we choose a 
smaller $r$. Then this guarantees that the resulting $Z(S)$ when RowCover 
fails is more sparse, which is desirable for us as described later. 
However since $|T| \le 1/r$, a smaller $r$ means a larger $T$ when the 
procedure succeeds, which costs more for G-queries in 
RowReduction. Thus, we should choose the minimum $r$ such that the query 
complexity for the case that RowCover keeps succeeding as long as the 
total cost does not exceed the total limit ($=O(\sqrt{N})$).

\subsection{Analysis of the Whole Algorithm}
Now we are ready to prove our first theorem:
\begin{theorem}\label{N_times_N}
The $M\times N$ OIP can be solved with a constant success probability by 
querying the blackbox oracle $O(\sqrt{N})$ times if $M = \mbox{poly}(N)$.
\end{theorem}
\begin{proof}
See Procedure 5 for the pseudocode of the algorithm ROIPS($S,Z$) (Robust 
OIP algorithm for Small $Z$). We call this procedure with $S = 
\{1,\ldots,M\}$ (we need this parameter  
since ROIPS is also used in the later algorithm) and the given matrix 
$Z$. As described in Sec.~3.1, we narrow the candidate set $S$ at lines 
2 and 3. If RowCover at line 2 succeeds, then $|S|$ 
is sufficiently reduced. Even if RowCover fails, $|S|$ is 
also reduced similarly if RowReduction at line 3 can find a $1$ by 
G-queries. Otherwise line 7 is executed where the current oracle 
looks like $f_2$ in Fig. 1. In this case, by finding a $1$ in the positions 
$\{1,\ldots,N\}\setminus{T}$ by the G-query at line 7, $|S|$ is reduced 
to $|S|\log^{4}{N}/N$, because we set $r = \log^{4}{N}/N$ at line 2. Since 
the original size of $S$ is $N^{c}$ for a constant $c$, line 7 is 
executed at most $c+1$ times. 

Note that the selection of the value of $r$ at line $2$ follows the rule 
described in Sec.~3.3: Since $r = \log^{4}{N}/N$, the size of $T$ at 
line 3 is at most $N/\log^{4}{N}$. This implies that the number of 
oracle calls at line 3 is $O(\log{N}\cdot \sqrt{N}/\log^{2}{N}) = 
O(\sqrt{N}/\log{N})$. Since line 3 is repeated at most $O(\log{N})$ 
times, the total number of oracle calls at line 3 is at most $O(\sqrt{N})$. 
Line 7 needs $O(\sqrt{N})$ oracle calls, but the number of its 
repetitions is $O(1)$ as mentioned above. Thus the total number of oracle 
calls is $O(\sqrt{N})$.  

Also by Lemma~1, the error probability of line 3 is at most $O(\log{N}/N)$. 
Since the number of repetitions is $O(\log{N})$, this error probability 
is obviously small enough. The error probability of line 7 is constant 
but again this is not harmful since it is repeated only $O(1)$ times, 
and thus the error probability can be made as small as it is needed at 
constant cost.
\end{proof}

\section{Algorithms for Large Candidate Sets}

\subsection{Overview of the Algorithm}
In this section, our $M\times N$ input  matrix $Z$ is large, i.e., $M$ is 
superpolynomial. We first observe how the previous algorithm, ROIPS, 
would work for such a large $Z$. Due to the rule given in Sec.~3.3, the 
value of $r$ at line 2 should be $\beta = 
\log{M}(\log\log{M})^2\log{N}/(2N)$. The calculation is not hard: Since 
we need $\log{M}$ repetitions for the main loop, we should assign 
roughly $\log\log{M}$ to $l$ of RowReduction for a sufficiently small 
error in each round. Then the cost of RowReduction will be 
$\sqrt{1/\beta}\cdot \log\log{M}$. Furthermore, we have to multiply 
the number of repetitions by $\log{M}$ factor, which gives us 
$\sqrt{N\log{M}/\log{N}}$, the desired complexity. Thus it would be nice 
if RowCover keeps succeeding. However, once RowCover fails, each column 
can still include as many as  $M\beta$~$1$'s which obviously needs too many repetitions 
of RowReduction at line 7 of ROIPS.               

Recall that the basic idea of ROIPS is to reduce the number of 
candidates in the candidate set $S$ by halving (the first phase) while 
the matrix is dense and to use the more direct method (the second phase) 
after the matrix becomes sufficiently sparse. If the original matrix is 
large, this strategy fails because, as mentioned above, the matrix does 
not become sufficiently sparse after the first phase. Now our idea is to 
introduce an "intermediate" procedure which reduces the number of the 
candidates  more efficiently than the first phase. For this purpose, we 
use RowReductionExpire\_MTGS, which tries to find a position of "1" in 
the oracle with multi-target Grover Search ($K > 1$ in Lemma~\ref{mtgs}) 
by assuming that the portion of such position, $K/N$, is sufficiently 
larger than $1/\beta$. If the assumption is indeed true then we apply 
RowReduction as before and moreover the number of G-queries in the main 
loop of RowReduction is repeated for a constant time of $\sqrt{N/K}$ 
on average. 

However, it is of course possible that the actual 
number of repetitions is far different from the expected value. That is 
why we limit the maximum number of oracle calls spent in G-queries by 
MAX\_QUERIES($N,M$), a properly adjusted number which depends on the 
size of the OIP matrix, and will be referred in the hereafter without 
its arguments for simplicity. If the value of COUNT gets this value, 
then the procedure expires (just stops) with no answer, but this 
probability is negligibly small by selecting  MAX\_QUERIES appropriately. 
Notice also that because of the failure of phase 1, it is  guaranteed 
that  the number of $1$'s in each column is "fairly" small, which in 
turn guarantees that the degree of row reduction is satisfactory for us. 
See  Procedure 8 for our new algorithm  ROIPL. 

Finally, when the assumption is false, RowReductionExpire\_MTGS finishes 
after $\log\log(\log{M}/\log{N})$ iterations of its main loop. In this 
case, we can prove that the matrix of the remaining candidates is very 
sparse and the number of its rows decreases exponentially by a single 
execution of RowReductionExpire\_MTGS. Thus one can achieve our upper bound 
also (details are given in the next section).

\subsection{Justification of the Algorithm}
One can see that in ROIPL, oracle calls take place only at lines 6 and 
11. As described in the previous overview, the total number of  
oracle calls in RowReduction at line 6 is $O(\sqrt{N\log{M}/\log{N}})$, 
and the whole execution of this part succesfully ends up with high probability. For 
the cost of line 11, we can prove the following lemma. 

\begin{lemma} The main loop (line 4 to 13) of ROIPL finishes with high 
probability before the value of COUNT reaches MAX\_QUERIES($N,M$).
\end{lemma}
\begin{proof} Note that there are two types of oracle calls in 
RowReductionExpire\_MTGS at lines 11. The first type, Type A, is when 
portion of "1" in the hidden oracle is at least 
$1/4(\log{|S|}/(N\log{N}))$, and the other type, Type B, is when the 
portion of "1" is smaller. Let $W = W_A+W_B$ be the expected number of 
oracle calls, where $W_A$ is the expected number of Type A calls and 
$W_B$, that of Type B calls. It is enough to prove that $W_A \le 
\frac{2}{3} \mbox{MAX\_QUERIES}$ and $W_B < \frac{1}{3} 
\mbox{MAX\_QUERIES}$. We defer the rigorous proofs in the Appendix and 
give instead the following more simple averaging argument on the bounds of 
$W_A$ and $W_B$. 

We first prove that $W_A \le \frac{2}{3}\mbox{MAX\_QUERIES}$. First, 
note that RowReductionExpire\_MTGS for Type A should require an $O(1)$ 
expected number of iterations of $\mbox{GQ}$, each of which requires 
$O(\sqrt{N\log{N}/\log{|S|}})$ queries. Now, since phase 1 has failed, 
the number of rows having a "1" at some position in $T=\{1..N\}$ is at 
most $\beta |S|$. Thus, after the above $O(\sqrt{N\log{N}/\log{|S|}})$ queries the 
number of candidates is reduced by a factor of $\beta = 
(\frac{1}{2})^{\log({1/\beta})}$. Therefore, intuitively, to reduce the 
number of candidates by half, the number of queries spent in $\mbox{GQ}(T)$ is  
$O(\frac{1}{\log({1/\beta})}\sqrt{N\log{N}/\log{|S|}})$.

Thus we have the following recurrence relation: 
\begin{eqnarray*}
W_A(|S|) \leq max(W_A(1), W_A(2), W_A(3), \cdots, W_A(|S|/2))+
O(\frac{1}{\log({1/\beta})}\sqrt{N\log{N}/\log{|S|}}), 
\end{eqnarray*}
where  $W_A(|S|)$ is the number of Type A queries to distinguish the candidate set $S$. 
Since ROIPL starts with $|S| = M$ and ends with $|S| \approx N^{10}$ (note that $\beta|S| > 
2$ if $|S| \approx N^{10}$), 
the above recurrence relation resolves to the following: 
\begin{eqnarray*}
W_A(M) & \leq & W_A(M/2)+ \sigma \frac{1}{\log({1/\beta})}\sqrt{N\log{N}/\log{M}} \\
& \le & \sigma \frac{\sqrt{N\log{N}}}{\log({1/\beta})}\left(\frac{1}{\sqrt{\log{M}}} +
  \frac{1}{\sqrt{\log(M/2)}} + \ldots +  
\frac{1}{\sqrt{10\log{N}}} \right) \\
& \le &  \sigma \frac{\sqrt{N\log{N}}}{\log({1/\beta})}\left(\frac{1}{\sqrt{\log{M}}} + \frac{1}{\sqrt{\log{M}-1}} + \ldots + 1 \right)\\
& \le & 2 \sigma \frac{\sqrt{N\log{N}}}{\log({1/\beta})} 
\sqrt{\log{M}} \le 2 \sigma \cdot \frac{\sqrt{N\log{M}\log{N}}}{\log{(1/\beta)}},
\end{eqnarray*}
where $\sigma$ is a sufficiently large constant. 
Therefore, the total number of queries is $O(\sqrt{N\log{M}/\log{N}})$ since  
$\log({1/\beta}) = \Omega(\log{N})$ if $M \le 2^{N^{d}}$. Note that if 
the above averaging argument is correct then  $|S|$ can be reduced into 
a constant by just repeating line~11. However, this is not exactly true 
for ROIPL since $|S|$ can only be reduced until becoming $\mbox{poly}(N)$ in order to 
obtain the desired number of query complexity (see the proof of Lemma~6 
in Appendix). Fortunately, in this case we can resort to ROIPS for 
identifying the hidden oracle out of $\mbox{poly(N)}$ candidates with 
just $O(\sqrt{N})$ queries as in line~16, and thus achieve a similar 
result with the averaging argument.

For technical details of ROIPL, note that $1/3\mbox{MAX\_QUERIES}$ is 
ten times the expected total number of queries supposing all queries are 
at line 11, i.e., the case with the biggest number of Type A queries. By 
Markov bound, the probability that the number of queries exceeds this 
amount is negligible (at most $1/10$). We summarize the property of 
RowReductionExpire\_MTGS in the following lemma which can be proven 
similarly as Lemma~\ref{rowreduct}.  
\begin{lemma}\label{mtgs}
The success probability and the number of oracle calls of the procedure \\
$\CALL{RowReductionExpire\_MTGS}{T,l,\mbox{COUNT},r}$ are $1-O(l/3^{l})$ and   
$l(O(\sqrt{1/r}) + l)$, respectively. Moreover, if there are more than 
$r$ fraction of $1$'s in the current oracle, then the average number of queries is 
$O(\sqrt{1/r}+l)$.
\end{lemma}

We next prove that $W_B < \frac{1}{3}\mbox{MAX\_QUERIES}$. In this 
case, MultiTargetGQ fails and therefore the density of "1" at every 
row of the candidates is less than $\gamma = 
\frac{1}{4}\log{|S|}/(N\log{N})$. Note that any two rows in $S''$ (the 
new S at the left-hand side of line 11) must be different, i.e.,  
we have to generate $|S''|$ different rows by using at most 
$\gamma N$~$1$'s for each row. Let $W$ be the number of rows in 
$S''$ which include at most $2\gamma N$~$1$'s. Then $|S''|-W$ rows include 
at least $2\gamma N$~$1$'s, and hence the number of such rows must be at 
most $|S|/2$. Thus we have $|S''|-W \le |S|/2$ and it follows that 
$$
|S''| \le 2W \le 2\sum_{k=0}^{\lambda= 
\lceil{2\gamma N}\rceil} {N \choose  
k}.
$$
The right-hand side is at most $2\cdot 2^{NH(\lambda/N)}$ (see e.g., 
\cite{CHLL97}, page 33), which is then bounded by $2|S|^{1/2}$ since $H(x) 
\approx x\log({1/x})$ for a small $x$. Thus, we have $|S''| \le 2 
|S|^{1/2}$. Hence, the number of candidates decreases doubly exponentially, 
which means we need only $O(\log(\log{M}/\log{N}))$ iterations of 
RowReductionExpire\_MTGS to reduce the number of the candidates from $M$ to 
$N^{10}$. Note that we let $l = \log\log(\log{M}/\log{N})$ at line 11 
and therefore the error probability of its single iteration is at most 
$O(1/\log(\log{M}/\log{N}))$. Considering the number of iterations 
mentioned above, this is enough to claim that $W_B < 
\frac{1}{3}\mbox{MAX\_QUERIES}$ (see Appendix for the proof in detail, where the 
actual bound of $W_B$ is shown to be much smaller). 
\end{proof}

Now here is our main theorem in this paper. 
\begin{theorem}\label{N_times_M_U}
The $M \times N$ OIP can be solved with a constant success probability
by querying the blackbox oracle $O(\sqrt{N 
\frac{\log{M}}{\log{N}}})$ times if $\mbox{poly}(N) \le M \le 2^{N^d}$ 
for some constant $d$ ($0 < d < 1$).     
\end{theorem}
\begin{proof}
The total number of oracle calls at line 6 is within the bound as 
described in Sec.~4.1 and the total number of oracle calls at line 11 is 
bounded by Lemma~4. As for the success probability,  we have already 
proved that there is no problem for the total success probability of 
line 6 (Sec.~4.1) and lines 11 (Lemma~4). Thus the theorem has 
been proved.  
\end{proof}


\subsection{OIP with $o(N)$ queries}

Next, we consider the case when $M > 2^{N^d}$. Note that when $M = 
2^{d'N}$, for a constant $d' \le 1$, the lower bound of the number of 
queries is $\Omega(N)$ instead of $\Omega(\sqrt{N\log{M}/\log{N}})$. 
Therefore, it is natural to expect that the number of queries exceeds 
our bound as $M$ approaches $2^{N}$. Indeed, when $2^{N^d} < M < 
2^{N/\log{N}}$, the number of queries of ROIPL is bigger than 
$O(\sqrt{N\log{M}/\log{N}})$ but still better than $O(N)$, as shown in 
the following theorem.           

\begin{theorem}\label{N_times_M_U1}
For $2^{N^d} \le M < 2^{N/\log{N}}$, the $M \times N$ OIP can be 
solved with a constant success probability by querying the blackbox oracle 
$O(\frac{\sqrt{N\log{N}\log{M}}}{\log({1/\beta})})$ times for $ 
 \beta = \mbox{~min}(\frac{\log{M}(\log\log{M})^2\log{N}}{2N},\frac{1}{4})$.
\end{theorem}
\begin{proof}
The algorithm is the same as ROIPL excepting the following: At line 1, 
we set $\beta$ as before if $M < 2^{N/\log^{3}{N}}$. Otherwise, i.e., if 
$2^{N/\log^{3}{N}} \le M \le 2^{N/\log{N}}$, we set $\beta = 1/4$. Then, 
we can use almost the same argument to prove the theorem, which may be 
omitted. 
\end{proof}

{\bf Remark~1~} Actually the query complexity of Theorem~3 changes 
smoothly from $O(\sqrt{N\log{M}/\log{N}})$ to $O(N/\log{N})$ and to 
$O(N)$ as $M$ changes from $2^{N^{d}}$ to $2^{N/\log^{3}{N}}$ and to 
$2^{N/\log{N}}$, respectively. When $M = 2^{N/\log{N}}$, the lower bound 
$\Omega(\sqrt{N\log{M}/\log{N}})$ in \cite{AIKMRY04} becomes 
$\Omega(N/\log{N})$. So it seems that our upper bound is worse than this 
lower bound by a factor of $\log{N}$. However, if $M$ is this large, 
then we can improve the lower bound to 
$\Omega(N/\sqrt{\log{N}\log\log{N}})$ and hence our upper bound is worse 
than the lower bound only by at most a factor of 
$O(\sqrt{\log{N}\log\log{N}})$ in this range (see Appendix).

\section{Concluding Remarks}
As mentioned above, our upper bound becomes trivial $O(N)$ when 
$M=2^{N/\log{N}}$, while for bigger $M$ \cite{BNRW03} has already 
given a nice robust algorithm which can be used for OIP with $O(N)$ queries. A challenging 
question is whether or not there exists an OIP algorithm whose upper 
bound is $o(N)$ for $M > 2^{N/\log{N}}$, say, for $M = 
2^{N/\log\log{N}}$. Even more challenging is to design an OIP algorithm 
which is optimal in the whole range of $M$. There are two possible 
scenarios: The one is that the lower bound becomes $\Omega(N)$ for some 
$M = 2^{o(N)}$. The other is that there is no such case, i.e., the bound 
is always $o(N)$ if $M = 2^{o(N)}$. At this moment, we do not have any 
conjecture about which scenario is more likely.  

\floatname{algorithm}{Procedure}
\begin{algorithm}
\caption{RowReduction($T,l$)}
\begin{algorithmic}[1]
\REQUIRE $T \subseteq \{1,\ldots,N\}$ and $l \in \mathcal{N}$ 
\FOR{$j \leftarrow 1 \TO l$} 
	\STATE $k \leftarrow \mbox{GQ}(T)$
	\IF{$\CALL{Majority}{k,\mbox{min}(l,\log{N}),f} = 1$} 
		\STATE \KEMBALI $\CALL{PositiveRow}{\{k\}, Z}$
	\ENDIF
\ENDFOR
\STATE \KEMBALI  $\{1,\ldots,M\} \setminus \CALL{PositiveRow}{T, Z}$
\end{algorithmic}
\end{algorithm}

\begin{algorithm}
\caption{RowCover($S,r$)}
\begin{algorithmic}[1]
\REQUIRE $S \subseteq \{1,\ldots,M\}$ and  $0 < r < 1$
\STATE $T \leftarrow \{ \}$
\STATE ${S}' \leftarrow S$
\WHILE{$\exists i \mbox{~s.t.~} |\mbox{PositiveRow}(\{i\}, Z(S'))| \ge
  r |S|$ and $|\mbox{PositiveRow}(T, Z(S))| < |S|/4$} 
\STATE $T \leftarrow T \cup \{i\}$
\STATE ${S}' \leftarrow S\setminus \mbox{PositiveRow}(T,Z(S))$
\ENDWHILE
\STATE \KEMBALI $T$ \COMMENT{{by one-sensitivity $|\mbox{PositiveRow}(T,Z(S))| < 3|S|/4$}}
\end{algorithmic}
\end{algorithm}

\begin{algorithm}
\caption{PositiveRow($T,Z$)}
\begin{algorithmic}
\STATE \KEMBALI $\{i|\ j \in T \mbox{~and~} Z(i,j) = 1\}$
\end{algorithmic}
\end{algorithm}

\floatname{algorithm}{Procedure}
\begin{algorithm}
\caption{Majority($k,l,f$)}
\begin{algorithmic}
\STATE \KEMBALI the majority of $60 l$ samples of $f(k)$  
if $k \neq \mbox{nill}$, else $0$.
\end{algorithmic}
\end{algorithm}

\begin{algorithm}
\caption{ROIPS($S,Z$)}
\begin{algorithmic}[1]
\REPEAT
	\STATE $T \GETS \CALL{RowCover}{S,\log^{4}{N}/N}$
	\STATE $S' \GETS S \cap \CALL{RowReduction}{T,\log{N}}$
	\IF{$|S'| \le \frac{3}{4}|S|$}
		\STATE $S \GETS S'$
	\ELSE
		\STATE $S \GETS S' \cap 
\CALL{RowReduction}{\{1,\ldots,N\}\setminus T, 1}$
	\ENDIF
\UNTIL{$|S| \le 1$}
\STATE $\KEMBALI{S}$
\end{algorithmic}
\end{algorithm}



\floatname{algorithm}{Procedure}
\begin{algorithm}
\caption{RowReductionExpire\_MTGS($T,l,\mbox{COUNT},r$)}
\begin{algorithmic}
\STATE \small{the same as RowReduction($T,l$) except that we add the
  folowing two: ($i$) 
  the number of queries is added to COUNT and the empty set is
  returned when COUNT exceeds $MAX\_QUERIES(N,M)$ (defined in ROIPL)} 
 ($ii$) For $r > 0$: $\mbox{GQ}(T)$ is replaced by
 $\mbox{MultiTargetGQ}(T,r)$, a G-query on $T$ assuming that there are
 more than $r$ fraction of $1$'s in the current oracle, and at line 7
 the set of all rows that have at most $r$ fraction of $1$'s is
 returned instead.  
\end{algorithmic}
\end{algorithm}

\begin{algorithm}[h]
\caption{ROIPL$(Z)$}
\begin{algorithmic}[1]
\REQUIRE $Z: M \times N$ 0-1 matrix and $\mbox{poly}(N) \le M \le 2^{N/\log{N}}$\\
\STATE $\beta \GETS \frac{\log{M}(\log\log{M})^2\log{N}}{2N};\ \ S = \{1,\ldots,M\}$
\STATE $\mbox{MAX\_QUERIES(N,M)} \GETS 45 \sigma\frac{\sqrt{N \log M 
\log N}}{\log{1/\beta}}$ \COMMENT{\small{$\sigma$: a constant factor of Robust 
Quantum Search in \cite{HMW03}}}\\
\STATE $\mbox{COUNT} \GETS 0$ \COMMENT{Increased in RowReductionExpire}\\
\REPEAT
 	\STATE $T \GETS \CALL{RowCover}{S,\beta}$
 	\STATE $S' \GETS S \cap \CALL{RowReduction}{T,\log{\log M}}$
 	\IF{$|S'| \le 3/4 |S|$}
 		\STATE $S \GETS S'$
 	\ELSE
 	        \STATE $S \GETS S'$
		\STATE $S \GETS S \cap 
\CALL{RowReductionExpire\_MTGS}{\{1 \ldots 
N\},\log\log(\frac{\log{M}}{\log{N}}),\mbox{COUNT},\frac{1}{4}\frac{\log{|S|}}{(N \log{N})})}$   
 	\ENDIF
\UNTIL{$|S|\le {N^{10}}$}
\STATE $Z' \GETS Z(S)$
\STATE relabel $S$ and $Z'$ so that the answer to OIP of $Z$ can be 
deduced from that of $Z'$
\STATE $\KEMBALI{\CALL{ROIPS}{S,Z'}}$
\end{algorithmic}
\end{algorithm}


\begin{appendix}
\section{Proof of Theorem~2}
Theorem~2 can be shown by proving Lemma~4, concluding 
that ROIPL succeeds to identify the blackbox oracle with constant 
probability using at most $O(\sqrt{N\log M/\log{N}})$ queries. Here, we 
provide its detailed proof by showing the following lemmas. Notice that $\sigma$ is 
the constant factor in Lemma~2 which can be computed from \cite{HMW03}.

\begin{lemma}\label{lemmaCandidateRowWithCounter} With high probability, 
the total number of Type A queries at line 11 in the whole rounds of 
ROIPL does not exceed $2/3 \cdot \mbox{MAX\_QUERIES(N,M)} =  
30\sigma\frac{\sqrt{N\log{M}\log{N}}}{\log({1/\beta})}$.    
\end{lemma}

\begin{lemma}\label{lemmaCandidateRow} With high probability, the total 
number of Type B queries at line 11 in the  whole 
rounds of ROIPL is less than  $1/3\cdot \mbox{MAX\_QUERIES(N,M)}= 
15\sigma\frac{\sqrt{N\log{M}\log{N}}}{\log({1/\beta})}$.     
\end{lemma}

Now it is left to prove the above two lemmas. 

{\bf Proof of Lemma~\ref{lemmaCandidateRowWithCounter}.}
Before proving Lemma~\ref{lemmaCandidateRowWithCounter},  we show the
following:  
\begin{lemma} RowReductionExpire\_MTGS at line 11 of 
ROIPL is executed for at most $m^* = \lceil{\frac{\log{M}-10\log{N}}{\log{1/\beta}}}\rceil$ times.  
\end{lemma}
\begin{proof}
RowReductionExpire\_MTGS at line 11 is executed when the first RowReduction at
line 6 cannot reduce $1/4$ fraction of the rows. Thus, finding a position of "1" reduces the 
number of candidates by a $\beta$ fraction. Thus, denoting the set of 
oracle candidates at round $k$ as  $S_k$, $|S_k|$ is at most 
$M\beta^{k}$. Therefore, it follows that RowReductionExpire is executed for at 
most $m^* = \lceil{\frac{\log{M}-10\log{N}}{\log{1/\beta}}}\rceil$ times.          
\end{proof}

Now, let us first bound the number of queries of Type A at 
RowReductionExpire\_MTGS at line 11. For this purpose, let $X_k$ and $X$ be 
the random variables denoting the number of  queries of 
the RowReductionExpire at round $k$ and the total number of queries of 
the RowReductionExpire in the whole rounds, respectively. Clearly, since 
for each trial of $GQ(T)$ the success probability is at least $2/3$, the 
average number of queries is:   
\begin{eqnarray*}
E[X] = \sum_{k=0}^{m^*} E[X_k] &\le& 
\sum_{k=0}^{m^*}\sum_{m=1}^{\infty}\sigma \cdot \frac{2}{3} \cdot
\frac{1}{3^{m-1}} \cdot m \cdot 
\sqrt{\frac{N\log{N}}{\log{|S_k|}}}\\   
&\le& \frac{3}{2}\sigma \cdot 
\sum_{k=0}^{m^*}\sqrt{\frac{N\log{N}}{\log{|S_k|}}}\\ 
&=& \frac{3}{2}\sigma \cdot 
\sqrt{N\log{N}} \sum_{k=0}^{m^*} 
1/\sqrt{\log{M} + k \log{\beta}} \\ 
&=& \frac{3}{2}\sigma \cdot 
\sqrt{N\log{N}} \sum_{k=0}^{m^*} 
1/\sqrt{10\log{N} + k \log{(1/\beta)}}\ \mbox{(reordering the 
summation)}\\  
&\le& \frac{3}{2}\sigma \cdot 
\sqrt{N\log{N}} 
\sum_{k=1}^{\log({\log{M}/\log{N}})-1} 
\frac{2^k\log{N}}{\log{(1/\beta)}}\cdot\frac{1}{\sqrt{2^k\log{N}}}\\  
&=& \frac{3}{2}\sigma \cdot 
\frac{\sqrt{N\log{N}\log{M}}}{\log{(1/\beta)}}. 
\end{eqnarray*}
Note that the fifth inequality is obtained from bounding the sum of 
terms whose values are between $\sqrt{\frac{N\log{N}}{2^k\log{N}}}$ and 
$\sqrt{\frac{N\log{N}}{2^{k+1}\log{N}}}$; there are at most $2^k 
\log{N}/\log{1/\beta}$ of them.

When $\mbox{poly}(N) \le M \le 2^{N^d}$, $\log{1/\beta} = 
\Omega(\log{N})$ and by Markov bound, $\mbox{Pr}[X \ge t \cdot E[X]] 
\le 1/t$, i.e., the probability that Stage~2 ends in failure is at 
most $\mbox{Pr}[X \ge 10 E[X]] \le 1/10$. This proves the lemma.      

{\bf Proof of Lemma~\ref{lemmaCandidateRow}.}
Since Type B queries are considered, the portion of "1" in the oracle is 
less than $1/4|S|/(N\log{N})$. Therefore if RowReductionExpire\_MTGS 
does not finish after $\log\log({\log{M}/\log{N}})$ repetitions, by 
Lemma~\ref{mtgs} this case can be detected  with probability at least 
$1-O(1/\log({\log{M}/\log{N}}))$. And fortunately, since $|S_k|$,  
the number of the candidate oracles at round $k$, is at most  
$2|S_{k-1}|^{1/2}$, this case happens only $\log({\log{M}/\log{N}})$ 
times in the whole course of the algorithm. 
Thus we have the following recurrence relation: 
\begin{eqnarray*}
W_B(|S_{k}|) \leq W_B(|S_{k+1}|) + O(\sqrt{\frac{N \log{N}}{\log{|S_k|}}}), 
\end{eqnarray*}
where  $W_B(|S_k|)$ is the number of Type B queries to distiguish the candidate set $S_k$. 
This resolves to 
$$
W_B(|S_0|) \le
\sum_{k=0}^{\log({\log{M}/\log{N}})} \sigma \sqrt{\frac{N 
\log{N}}{\log{|S_k|}}}\cdot \log\log({\log{M}/\log{N}}) \le 
3\sigma \sqrt{N} \log\log({\log{M}/\log{N}}),   
$$
which is much smaller than $1/3\cdot \mbox{MAX\_QUERIES}$ since 
$\log\log{x} \le \sqrt{x}$ for $x \ge 1$ and $\log{(1/\beta)} =  
\Omega(\log{N})$ for $M \le 2^{N^d}$. As can be seen in the above 
inequality, the number of queries at the last rounds, namely, when $|S_k| = 
poly(N)$, is the dominant factor because $|S_k|$ decreases  doubly 
exponentially. This concludes the proof.  

\section{Slightly Better Lower Bounds for OIP}
Here, we will show that for $2^{N^d} < M \le 2^{N/\log{N}}$ ROIPL is 
only $\sqrt{\log{N}\log\log{N}}$ worse than the query-optimal algorithm. The 
following theorem is by \cite{AIKMRY04}.    

\begin{theorem} There exists an OIP whose query complexity is 
$\Omega(\sqrt{N\log{M}/\log{N}})$.   
\end{theorem} 

By a simple argument, indeed the above theorem can be restated more 
accurately as follows.    

\begin{theorem}
There exists an OIP whose query complexity is 
$\Omega(\sqrt{(N-\tilde{k})(\tilde{k}+1)})$ when the number of 
candidates $M$ satisfies    
$$
{N \choose \tilde{k}-1} + {N \choose \tilde{k}} \le M \le {N \choose 
\tilde{k}} + {N \choose \tilde{k}+1}.  
$$
\end{theorem}
\begin{proof}
Similar to the proof in Theorem~2 in \cite{AIKMRY04}. In fact, the proof 
of Theorem~2 of \cite{AIKMRY04} already achieved the above lower bound 
but there $\Omega(\log{M}/\log{N})$ is substituted for $\tilde{k}$ which 
is not done here because the substitution can weaken the statement.  
\end{proof}

{\bf Remark~2~}A similar but weaker lower bound can be found in 
\cite{FGGS99} where it is shown that the lower bound for OIP with the 
number of candidates $M$ is $\tilde{k}$ such that $\tilde{k}$ is the 
smallest integer satisfying $M \le \sum_{l=0}^{\tilde{k}}{N \choose l}$. 

Now, we can state the following lemma.
\begin{lemma}
For $M \le 2^{N/\log{N}}$, ROIPL is at most $O(\sqrt{\log{N}\log\log{N}})$ 
worse than the optimal algorithm. 
\end{lemma}
\begin{proof} For $M = 2^{N/\log{N}}$, we can take $\tilde{k}/N 
\log({N/\tilde{k}}) = 1/\log{N}$ since ${N \choose \lambda{N}} =  
2^{(1-o(1))N{H(\lambda)}}$ (see, e.g., \cite{CHLL97}, page 33) where 
here, $H(x) \approx x\log({1/x})$ for a small $x$ . By the previous 
theorem, there exists an OIP whose query complexity is 
$\Omega(N/\sqrt{\log{N}\log\log{N}})$ while by Theorem~3 the query 
complexity of ROIPL is only $O(N)$.   
\end{proof}
\end{appendix}
\end{document}